\documentclass{article}

\usepackage[pdftex]{graphicx, color}
\usepackage[colorlinks=true, linkcolor=black, filecolor=black, urlcolor=black, citecolor=black]{hyperref}
\usepackage{amsmath,graphicx,amssymb,amsfonts,bm,color,url,mathtools,cite}
\allowdisplaybreaks[4]

\usepackage{authblk}
\usepackage{indentfirst}
\usepackage{geometry}
\allowdisplaybreaks[2]

\def\lr{\bm{\lambda}_{\mathrm{R}}}
\def\sr{s_{\mathrm{R}}}

\makeatletter
\long\def\@makecaption#1#2{%
  \normalsize
  \vskip\abovecaptionskip
  \sbox\@tempboxa{#1: #2}%
  \ifdim \wd\@tempboxa >\hsize
    #1: #2\par
  \else
    \global \@minipagefalse
    \hb@xt@\hsize{\hfil\box\@tempboxa\hfil}%
  \fi
  \vskip\belowcaptionskip}
\makeatother

\title{DisC-VC: Disentangled and $F_0$-Controllable Neural Voice Conversion}
\author[1]{Chihiro Watanabe\thanks{chihiro.watanabe.xz@hco.ntt.co.jp}} 
\author[1]{Hirokazu Kameoka\thanks{This work was supported by JST CREST Grant Number JPMJCR19A3, Japan.}}
\affil[1]{{\normalsize NTT Communication Science Laboratories, 3-1, Morinosato Wakamiya, Atsugi-shi, Kanagawa Pref. 243-0198 Japan}}
\date{}

\begin{document}
\maketitle

\begin{abstract}
Voice conversion is a task to convert a non-linguistic feature of a given utterance. Since naturalness of speech strongly depends on its pitch pattern, in some applications, it would be desirable to keep the original rise/fall pitch pattern while changing the speaker identity. Some of the existing methods address this problem by either using a source-filter model or developing a neural network that takes an $F_0$ pattern as input to the model. Although the latter approach can achieve relatively high sound quality compared to the former one, there is no consideration for discrepancy between the target and generated $F_0$ patterns in its training process. In this paper, we propose a new variational-autoencoder-based voice conversion model accompanied by an auxiliary network, which ensures that the conversion result correctly reflects the specified $F_0$/timbre information. We show the effectiveness of the proposed method by objective and subjective evaluations. 

\end{abstract}

\section{Introduction}
\label{sec:introduction}

Voice conversion (VC) is a task to convert a non-linguistic feature of a given utterance and it can be used for various applications \cite{Mohammadi17} including personalizing or changing emotion of text-to-speech systems \cite{Kain98, Inanoglu09, Robinson19}. 
In this paper, we focus on the conversion of $F_0$ pattern and timbre (i.e., speaker identity). 

A basic VC framework consists of feature extraction from a speech signal, feature conversion, and speech synthesis based on a converted feature. With regard to feature conversion, some of the existing methods require parallel data set (e.g., Gaussian-mixture-model-based \cite{Kain98b, Stylianou98, Toda07}, deep-neural-network-based \cite{Desai09, Desai10}, and sequence-to-sequence-based \cite{Ramos16, Zhang19} methods), while others do not (e.g., recognition-synthesis-based \cite{Huang21, Lin21}, phonetic-posteriorgram-based \cite{Sun16}, variational-autoencoder (VAE) \cite{Kingma14}-based \cite{Kameoka19, Qian19, Qian20}, generative-adversarial-network-based \cite{Hsu17, Kaneko18, Kaneko19}, flow-based \cite{Serra19, Merritt22}, and score-based \cite{Kameoka20, Popov22} methods). As for feature extraction and speech synthesis, the existing methods can be classified roughly into two approaches. 
One is the parametric-vocoder-based (PV) approach \cite{Stylianou98, Desai09, Kameoka19}, where we first apply timbre conversion based on mel-cepstral coefficients (MCCs) and $F_0$ conversion (e.g., linear transformation) separately and then synthesize a waveform with a parametric vocoder such as WORLD \cite{Morise16}. 
The other is the neural-vocoder-based (NV) approach \cite{Biadsy19, Kameoka19s, Qian19}, where we use a neural network for speech synthesis. While early neural vocoders have required spectral envelope- and excitation-related features as input \cite{Hayashi17, Tamamori17}, the recent mainstream is to synthesize a waveform from a mel-spectrogram \cite{Kong20, Yamamoto20}. 

In some applications such as online communication, it would be desirable to keep the original rise/fall pitch pattern, since it strongly affects the nuance of speech. Inappropriate modification of the relative pitch pattern would cause miscommunication in regard to the speaker's intent and/or emotion. 
One way to securely keep the relative pitch pattern of input speech would be to take the PV approach, which allows us to explicitly specify the $F_0$ pattern of the converted speech. However, one disadvantage is that it cannot generally achieve as high sound quality as the NV approach. 
Recently, some $F_0$-controllable NV methods have been proposed, with which we can specify both $F_0$ and timbre conditionings explicitly \cite{Qian20b, Liu21, Kim22}. 
However, even with these models, a generated speech does not necessarily have the same $F_0$ pattern as the input conditioning. This is because the relationship between the specified $F_0$ pattern and the conversion result has not been considered in the training criteria, and thus the decoder may disregard the specified $F_0$ pattern and not reproduce it faithfully in the reconstruction process. 

In this paper, we develop a disentangled and $F_0$-controllable neural VC (DisC-VC). 
DisC consists of the VAE-based generator and auxiliary network. The auxiliary network, which is composed of $F_0$ extractor and speaker classifier, ensures that the conversion result correctly reflects the specified $F_0$/timbre information. We show the effectiveness of DisC by both objective and subjective evaluations. 


\section{Proposed Method}
\label{sec:method}

\subsection{Preprocessing of training samples}
\label{sec:preprocessing}

\begin{figure}[t]
  \centering
  \includegraphics[width=0.8\linewidth]{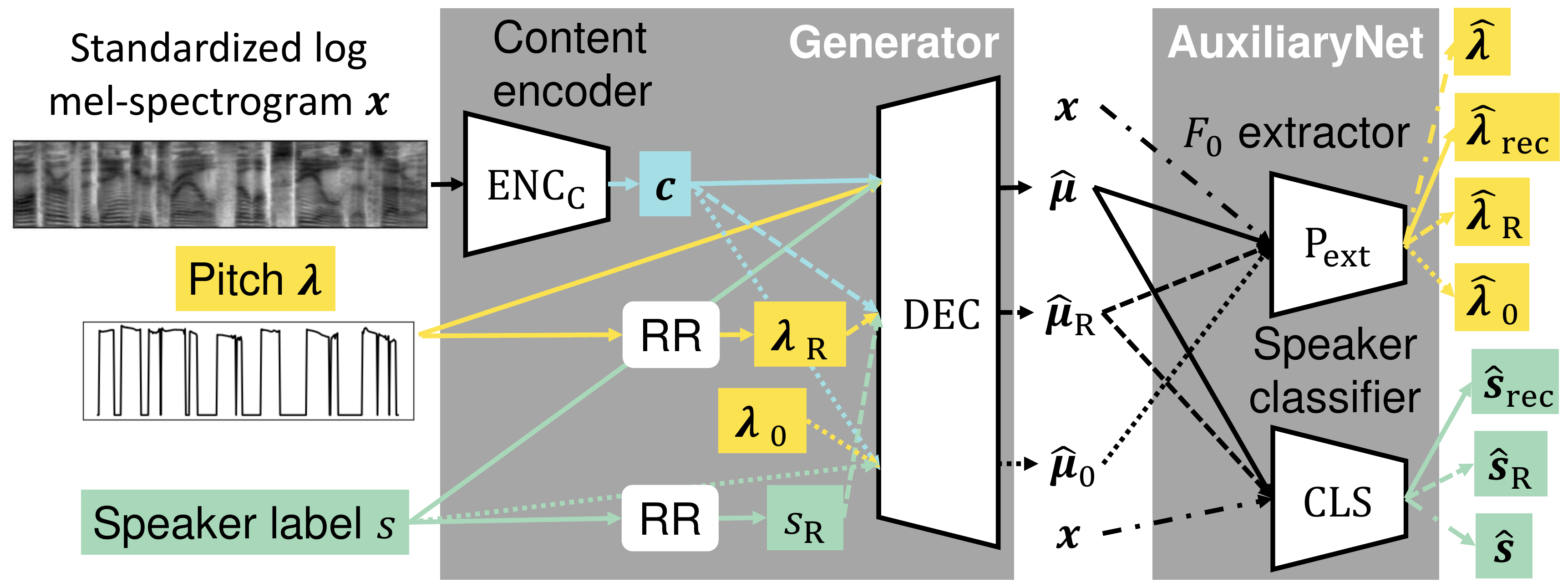}
  \includegraphics[width=0.8\linewidth]{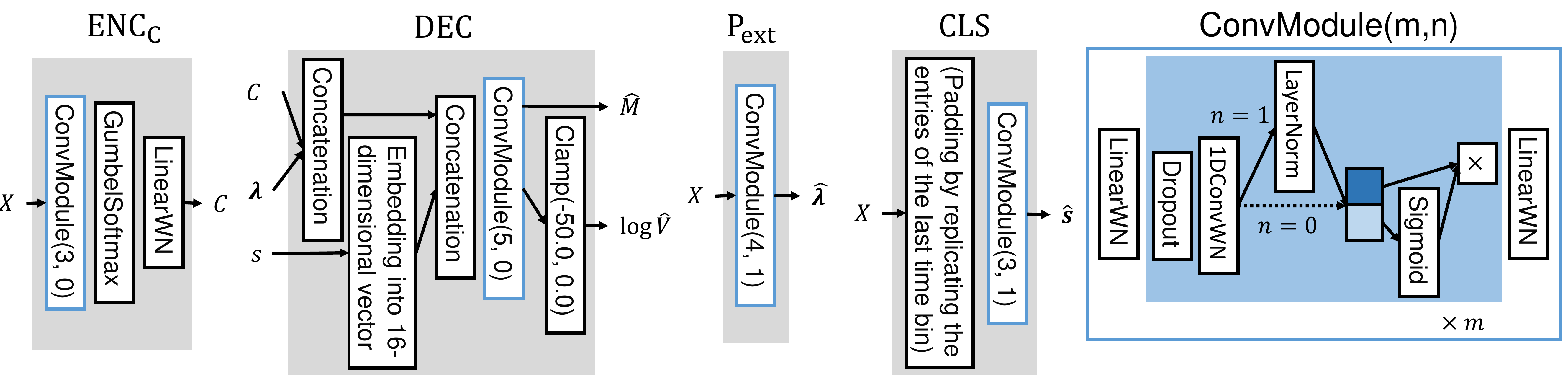}
  \caption{The network architecture of the DisC-VC. ``GumbelSoftmax'' means sampling from the Gumbel-Softmax distribution \cite{Jang17}. ``LinearWN'' and ``1DConvWN,'' respectively, are linear and 1D convolution layers with weight normalization \cite{Salimans16}. $\mathrm{Clamp}(a, b)$ is a function that clamps input into the range $[a, b]$. ``LayerNorm'' means layer normalization \cite{Ba16}.}
\label{fig:architecture}
\end{figure}

Figure \ref{fig:architecture} shows the architecture of the DisC model. During the training phase, we use the standardized log mel-spectrogram $\bm{x} \in \mathbb{R}^{F \times T}$, the log $F_0$ pattern $\bm{\lambda} \in \mathbb{R}^T$, and the speaker index $s \in \{ 1, 2, \dots, S \}$ as the input features, where $T$, $F$, and $S$ indicate the numbers of time and frequency bins and speakers, respectively. To define $\bm{x}$, we first compute the log mel-spectrogram $\bm{x}^{(0)}$. Then, each entry of the column vectors of $\bm{x}^{(0)}$ is standardized according to the mean and standard deviation for all the time bins and training samples. To compute $\bm{\lambda}$, we extract the $F_0$ pattern from an input speech signal with the WORLD $F_0$ estimator \cite{Morise16} and apply logarithmic transformation to its non-zero entries. 

\subsection{Generator}
\label{sec:generator}

The generator consists of content encoder $\mathrm{ENC}_{\mathrm{C}}$ and decoder $\mathrm{DEC}$. To reflect the discrete nature of the phonemic units of speech, we adopt the encoder of Vector Quantised-Variational AutoEncoder or VQ-VAE \cite{Oord17} as $\mathrm{ENC}_{\mathrm{C}}$. In the generator, we first extract the content feature $\bm{c} = \mathrm{ENC}_{\mathrm{C}} (\bm{x}) \in \mathbb{R}^{N_{\mathrm{C}} \times T}$ from $\bm{x}$, where we set $N_{\mathrm{C}} = 16$ in the experiment. 
Then, the decoder reconstructs the mean and variance of the log mel-spectrogram $(\hat{\bm{\mu}}, \hat{\bm{\sigma}}) = \mathrm{DEC} (\bm{c}, \bm{\lambda}, s)$, where $\hat{\bm{\mu}}, \hat{\bm{\sigma}} \in \mathbb{R}^{F \times T}$. 

\subsection{Training objective}

\subsubsection{Reconstruction error}
\label{sec:l_like}

The first training objective is the reconstruction error of the generator. Let $q_{\bm{\phi}} (\bm{c}|\bm{x})$ be a probability density function (PDF) of content feature $\bm{c}$ conditioned on log mel-spectrogram $\bm{x}$, based on which the content encoder with parameter $\bm{\phi}$ generates the content feature. Let $p_{\bm{\theta}} (\bm{x}|\bm{c}, \bm{\lambda}, s)$ be a PDF of $\bm{x}$ conditioned on a set of features $(\bm{\lambda}, s)$, based on which the decoder with parameter $\bm{\theta}$ generates the output mel-spectrogram. We assume the prior distribution of $\bm{c}$ to be a discrete uniform distribution independent of the $F_0$/timbre features (i.e., $p(\bm{c} | \bm{\lambda}, s) = p(\bm{c})$). 
By using Jensen's inequality, the negative log marginal distribution can be upper bounded by
\begin{align}
\label{eq:elbo}
-\log p_{\bm{\theta}} (\bm{x} | \bm{\lambda}, s) 
\leq D - \mathbb{E}_{\bm{c} \sim q_{\bm{\phi}} (\bm{c}|\bm{x})} [\log p_{\bm{\theta}} (\bm{x}|\bm{c}, \bm{\lambda}, s)],
\end{align}
where $D \coloneqq D_{\mathrm{KL}} [q_{\bm{\phi}} (\bm{c}|\bm{x}) || p (\bm{c})]$ and $D_{\mathrm{KL}} (q||p)$ indicates the KL divergence of $q$ from $p$. It can be shown that the difference between the right- and left-hand sides of (\ref{eq:elbo}) is equal to $D_{\mathrm{KL}} (q_{\bm{\phi}} (\bm{c}|\bm{x})||p_{\bm{\theta}} (\bm{c}|\bm{x}, \bm{\lambda}, s))$, which is minimized when $q_{\bm{\phi}} (\bm{c}|\bm{x}) = p_{\bm{\theta}} (\bm{c}|\bm{x}, \bm{\lambda}, s)$ (i.e., the content feature $\bm{c}$ extracted by the encoder is independent of $(\bm{\lambda}, s)$, given $\bm{x}$). When $p(\bm{c})$ is a discrete uniform distribution, $D$ becomes constant \cite{Oord17}. 
By assuming $p_{\bm{\theta}} (\bm{x}|\bm{c}, \bm{\lambda}, s)$ to be a Gaussian distribution with mean $\hat{\bm{\mu}}$ and variance $\hat{\bm{\sigma}}$, where all the entries are mutually independent, the second term in (\ref{eq:elbo}) can be interpreted as the reconstruction error of $\bm{x}$. Given training samples of $(\bm{x}, \bm{\lambda}, s)$, the actual training criterion $\mathcal{L}_{\mathrm{like}}$ will be the expectation of the right-hand side of (\ref{eq:elbo}) over $(\bm{x}, \bm{\lambda}, s)$. 

\subsubsection{Negative mutual information}

As we later show experimentally in Section \ref{sec:experiment}, even if the conversion model is conditioned on specified $F_0$/timbre information $(\lr, \sr)$, the converted speech does not always reflect them sufficiently with a model trained only based on the reconstruction loss $\mathcal{L}_{\mathrm{like}}$, since the decoder may disregard $(\lr, \sr)$ and only use $\bm{c}$ to reconstruct the log mel-spectrogram. Therefore, as the second training objective, given $\bm{c} \sim q_{\bm{\phi}} (\bm{c}|\bm{x})$ and $\sr$, we consider the negative mutual information $\mathcal{I}_{\mathrm{P}} (\bm{\theta})$ between output log mel-spectrogram $\bm{x} \sim p_{\bm{\theta}} (\bm{x}|\bm{c}, \lr, \sr)$ and $\lr$, which is given by
\begin{align}
\label{eq:mi_f}
&\mathcal{I}_{\mathrm{P}} (\bm{\theta}) = -\mathbb{E}_{\bm{x} \sim p_{\bm{\theta}} (\bm{x}|\bm{c}, \lr, \sr), \lr' \sim p(\lr'|\bm{x})} [\log p(\lr'|\bm{x})] + H_{\mathrm{P}}, 
\end{align}
where $H_{\mathrm{P}} \coloneqq \mathbb{E}_{\lr \sim p(\lr)} [\log p(\lr)]$ is a constant that does not depend on $\bm{\theta}$. 
By using the fact that a KL divergence is non-negative and the law of total expectation, the expectation of the first term in (\ref{eq:mi_f}) over $\lr$ can be upper bounded by
\begin{align}
\label{eq:logr_f}
&-\mathbb{E}_{\lr \sim p(\lr), \bm{x} \sim p_{\bm{\theta}} (\bm{x}|\bm{c}, \lr, \sr), \lr' \sim p(\lr'|\bm{x})} [\log p(\lr'|\bm{x})] \nonumber \\
&\leq -\mathbb{E}_{\lr \sim p(\lr), \bm{x} \sim p_{\bm{\theta}} (\bm{x}|\bm{c}, \lr, \sr), \lr' \sim p(\lr'|\bm{x})} [\log q (\lr'|\bm{x})] \nonumber \\
&= -\mathbb{E}_{\lr \sim p(\lr), \bm{x} \sim p_{\bm{\theta}} (\bm{x}|\bm{c}, \lr, \sr)} [\log q (\lr|\bm{x})] 
\eqqcolon R_{\mathrm{P}}, 
\end{align}
where $q (\lr'|\bm{x})$ is an arbitrary conditional PDF. 
The equality holds if $q (\lr'|\bm{x}) = p(\lr'|\bm{x})$. Therefore, assuming $q (\lr'|\bm{x})$ in some parametric form, the left-hand side of (\ref{eq:logr_f}) can be decreased by alternately decreasing $R_{\mathrm{P}}$ with respect to $(\bm{\phi}, \bm{\theta})$ and $q$. 
Specifically, for all $i \in \{ 1, \dots, T \}$, we represent $q ((\lambda_{\mathrm{R}}')_i|\bm{x})$ as the Laplace distribution with location parameter $(\hat{\lambda}_{\mathrm{R}})_i$ and scale parameter of one, where $\hat{\bm{\lambda}}_{\mathrm{R}} = \mathrm{P}_{\mathrm{ext}} (\hat{\bm{\mu}}_{\mathrm{R}})$ is an output of an auxiliary neural network (i.e., $F_0$ extractor) $\mathrm{P}_{\mathrm{ext}}: \mathbb{R}^{F \times T} \mapsto \mathbb{R}^T$ and $(\hat{\bm{\mu}}_{\mathrm{R}}, \hat{\bm{\sigma}}_{\mathrm{R}}) = \mathrm{DEC} (C, \lr, \sr)$,
and consider training its parameters $\bm{\psi}_{\mathrm{P}}$ along with $\bm{\phi}$ and $\bm{\theta}$. Hence, $\mathrm{P}_{\mathrm{ext}}$ is trained to estimate the log $F_0$ pattern from an input log mel-spectrogram. 

Given $\bm{c}$ and $\lr$, we also consider using the negative mutual information $\mathcal{I}_{\mathrm{T}} (\bm{\theta})$ between $\bm{x}$ and $\sr$ as a training criterion. 
As above, the expectation of $\mathcal{I}_{\mathrm{T}} (\bm{\theta})$ except for a constant term over $\sr$ can be upper bounded by $R_{\mathrm{T}} \coloneqq -\mathbb{E}_{\sr \sim p(\sr), \bm{x} \sim p_{\bm{\theta}} (\bm{x}|\bm{c}, \lr, \sr)} [\log r (\sr|\bm{x})]$, where $r(\sr|\bm{x})$ is an arbitrary categorical distribution and the exact bound is achieved when $r(\sr|\bm{x}) = p(\sr|\bm{x})$. Thus, it can be decreased by alternately decreasing $R_{\mathrm{T}}$ with respect to $r$ and $(\bm{\phi}, \bm{\theta})$. Here, we assume $r(\sr|\bm{x})$ to be a categorical distribution with each probability given as an output of an auxiliary neural network (i.e., speaker classifier) $\mathrm{CLS}: \mathbb{R}^{F \times T} \mapsto \mathbb{R}^{T_{\mathrm{S}}}$ \footnote{$T_{\mathrm{S}}$ is an arbitrary natural number. As shown in Figure \ref{fig:architecture}, $\mathrm{CLS}$ contains convolutional layers and thus $T_{\mathrm{S}}$ depends on $T$ in the experiment.}, and consider training its parameters $\bm{\psi}_{\mathrm{T}}$ along with $\bm{\phi}$, $\bm{\theta}$, and $\bm{\psi}_{\mathrm{P}}$. Hence, $\mathrm{CLS}$ is trained to detect the speaker from an input log mel-spectrogram. 

The criteria $R_{\mathrm{P}}$ and $R_{\mathrm{T}}$ are defined under specific $(\bm{c}, \sr)$ and $(\bm{c}, \lr)$, respectively. 
Since $\bm{c}$, $\sr$, and $\lr$ are determined stochastically according to the training examples, the actual criteria used for training will be the expectations of $R_{\mathrm{P}}$ and $R_{\mathrm{T}}$ with respect to $\bm{c}$, $\sr$, and $\lr$. 
Let $\mathcal{L}_{\mathrm{P}} \coloneqq\mathbb{E}_{\bm{x}' \sim p (\bm{x}'), \bm{c} \sim q_{\bm{\phi}} (\bm{c}|\bm{x}'), \sr \sim p(\sr)} [R_{\mathrm{P}}]$ and $\mathcal{L}_{\mathrm{T}} \coloneqq\mathbb{E}_{\bm{x}' \sim p (\bm{x}'), \bm{c} \sim q_{\bm{\phi}} (\bm{c}|\bm{x}'), \lr \sim p(\lr)} [R_{\mathrm{T}}]$, where $\bm{x}' \sim p (\bm{x}')$ indicates the log mel-spectrogram of a randomly drawn training sample. 
The process $\lr \sim p(\lr)$ can be implemented as random resampling (RR), which converts an input log $F_0$ pattern $\bm{\lambda}$ into random one $\lr$. In the experiment, we defined the RR of $F_0$ by $(\lambda_{\mathrm{R}})_i = \lambda_i + \beta$ with $\beta \sim \mathrm{Uniform} ([0.3, 3))$, if $\lambda_i \neq 0$, and $(\lambda_{\mathrm{R}})_i = \lambda_i$, otherwise. 
We also defined the RR of speaker index by generating $\sr$ from a discrete uniform distribution on $\{ 1, \dots, S \}$. 
The idea of RR has also been adopted in SpeechFlow \cite{Qian20}, however, there has been no consideration for the discrepancy between the specified $F_0$/timbre information and those of the conversion result. 

The actual flow of computing $\mathcal{L}_{\mathrm{P}}$ and $\mathcal{L}_{\mathrm{T}}$ can be explained as follows. For each training sample $\bm{x}'$, the decoder generates $\bm{x}$ using $\bm{c} = \mathrm{ENC}_{\mathrm{C}} (\bm{x}')$ along with $\lr$ and $\sr$ randomly drawn from predefined distributions $p(\lr)$ and $p(\sr)$, respectively. The auxiliary network then extracts $\hat{\bm{\lambda}}_{\mathrm{R}}$ and $\hat{\bm{s}}_{\mathrm{R}}$ from $\bm{x}$ and checks whether they are consistent with $\lr$ and $\sr$. 
In addition, to better express unvoiced and voiced speech differently, we define $\mathcal{L}_{\mathrm{P0}}$ for a zero vector $\bm{\lambda}_0 \in \mathbb{R}^T$ and $\hat{\bm{\lambda}}_0 = \mathrm{P}_{\mathrm{ext}} (\hat{\bm{\mu}}_0)$ as a training criterion, where $(\hat{\bm{\mu}}_0, \hat{\bm{\sigma}}_0) = \mathrm{DEC} (C, \bm{\lambda}_0, s)$. We also define $\mathcal{L}_{\mathrm{P1}}$ for $\bm{\lambda}$ and $\hat{\bm{\lambda}} = \mathrm{P}_{\mathrm{ext}} (\bm{x})$, $\mathcal{L}_{\mathrm{P2}}$ for $\bm{\lambda}$ and $\hat{\bm{\lambda}}_{\mathrm{rec}} = \mathrm{P}_{\mathrm{ext}} (\hat{\bm{\mu}})$, $\mathcal{L}_{\mathrm{T1}}$ for $s$ and $\hat{\bm{s}} = \mathrm{CLS} (\bm{x})$, and $\mathcal{L}_{\mathrm{T2}}$ for $s$ and $\hat{\bm{s}}_{\mathrm{rec}} = \mathrm{CLS} (\hat{\bm{\mu}})$. 

We train the entire network to minimize $\mathcal{L} \coloneqq \eta \mathcal{L}_{\mathrm{like}} + \eta_{\mathrm{P}} (\mathcal{L}_{\mathrm{P}} + \mathcal{L}_{\mathrm{P0}} + \mathcal{L}_{\mathrm{P1}} + \mathcal{L}_{\mathrm{P2}}) + \eta_{\mathrm{T}} (\mathcal{L}_{\mathrm{T}} + \frac{1}{2} \mathcal{L}_{\mathrm{T1}} + \frac{1}{2} \mathcal{L}_{\mathrm{T2}})$ based on back-propagation. 
In the experiment, we set $\eta = 1/FT$, $\eta_{\mathrm{P}} = 1/4T$, and $\eta_{\mathrm{T}} = 1/2T_{\mathrm{S}}$, and replaced expectation over $\bm{x}'$ in computing each loss with a sample mean of a given mini-batch, which consists of $16$ randomly selected training samples. A training sample was defined as $\{ (\bm{x}_s, \bm{\lambda}_s, s) | s = 1, \dots, S \}$, where $\bm{x}_s$ and $\bm{\lambda}_s$ were a log mel-spectrogram and a log $F_0$ pattern of the $s$th speaker, respectively. It must be noted that the criterion $\mathcal{L}$ can be used not only for training the generator considered in this study, but also for training other existing generator networks that allow $F_0$/timbre conditioning (e.g., SpeechFlow \cite{Qian20}). 

\subsection{Conversion Process}
\label{sec:vc}

By using the trained DisC model, we can apply the following types of conversion to a given test sample $(\bm{x}, \bm{\lambda}, s)$: (1) change log $F_0$ pattern by a given value $\beta \in \mathbb{R}$, (2) change log $F_0$ pattern from speaker $s$ to $\check{s}_{\mathrm{P}}$, and (3) change timbre from speaker $s$ to $\check{s}_{\mathrm{T}}$. 
By defining $\beta = 0$, $\check{s}_{\mathrm{P}} = s$, or $\check{s}_{\mathrm{T}} = s$, we can also perform each type of conversion individually. 

The target log $F_0$ pattern $\check{\bm{\lambda}} \in \mathbb{R}^T$ is defined as follows. Let $\mu_{\mathrm{P}} (s)$ and $\sigma_{\mathrm{P}} (s)$, respectively, be the sample mean and sample standard deviation of the log $F_0$ pattern (except for the zero entries) of the $s$th speaker in the training data set. 
For all $i \in \{ 1, \dots, T \}$, we define $\check{\lambda}_i = [\sigma_{\mathrm{P}} (\check{s}_{\mathrm{P}})/\sigma_{\mathrm{P}} (s)] [\lambda_i - \mu_{\mathrm{P}} (s)] + \mu_{\mathrm{P}} (\check{s}_{\mathrm{P}}) + \beta$, if $\lambda_i \neq 0$, and $\check{\lambda}_i = 0$, otherwise. 

We obtain $(\hat{\bm{\mu}}, \hat{\bm{\sigma}}) = \mathrm{DEC} (\mathrm{ENC}_{\mathrm{C}} (\bm{x}), \check{\bm{\lambda}}, \check{s}_{\mathrm{T}})$ and reconstruct a waveform from $\hat{\bm{\mu}}$ with a neural vocoder. In the experiment, we used Parallel WaveGAN \cite{Yamamoto20}. 


\section{Experiment}
\label{sec:experiment}

\subsection{Experimental settings}

By using the CMU Arctic speech databases \cite{Kominek04}, we trained the DisC model with the $1,000$ training samples per speaker for $5,000$ epochs, applied $F_0$ and/or timbre conversion to the $132$ test samples per speaker, and compared the results with those obtained with the following baseline methods. 

DisC-NoAux is DisC with no auxiliary network, and it is trained by minimizing $\mathcal{L}_{\mathrm{like}}$ in Section \ref{sec:l_like}. ACVAE-C \cite{Kameoka19} is the original ACVAE that takes MCCs as input and reconstructs a waveform from converted MCCs and $F_0$ pattern by using a WORLD vocoder \cite{Morise16}. 
ACVAE-S \cite{Kameoka19s} is the extended ACVAE that takes mel-spectrogram as input and reconstructs a waveform from a converted mel-spectrogram by using a neural vocoder. 
For reference, we also used WORLD vocoder \cite{Morise16} itself for $F_0$-only conversion by using the MCCs of an input signal and $\check{\bm{\lambda}}$ given in Section \ref{sec:vc}. We also tried $F_0$ and timbre conversion with BNE-Seq2seqMoL (BNE) \cite{Liu21} as an example of $F_0$-controllable methods. It must be noted that we cannot compare its result with those of the other methods as it is, since we used the pretrained synthesis model provided by the authors of \cite{Liu21}, which was trained with a different data set than the CMU Arctic one. 

To compare these methods, we tried \textbf{task P} to change log $F_0$ pattern by $\beta \in \{ \log (1.5), -\log (1.5) \}$, \textbf{task T} to change timbre to that of a target speaker, and \textbf{task PT} to change both $F_0$ pattern and timbre to those of a target speaker. Since only simultaneous conversion of $F_0$ pattern and timbre is possible with ACVAE-S, we use the same results in tasks T and PT for this method. Before evaluation, we trimmed the leading and trailing silent sections of all the target and converted samples. 

\subsection{Objective evaluation}

To evaluate $F_0$ and timbre conversion performance, we adopted the $F_0$ error $\Delta F_0$ (i.e., the root mean squared error between two log $F_0$ patterns) and the mel-cepstral distortion (MCD), respectively. 
Since target and converted speech signals are not necessarily aligned in the time direction, we first applied endpoint-free Dynamic Time Warping  to a given pair of log $F_0$ patterns or MCCs and then computed $\Delta F_0$ and the MCD with the optimally aligned sequences. 
For all the tasks, we computed $\Delta F_0$ between the target log $F_0$ pattern $\check{\bm{\lambda}}$ and that of the converted speech. 
We evaluated the MCD between the samples before and after conversion for task P and those between the target and converted samples for tasks T and PT. 

Tables \ref{tab:objective_f}, \ref{tab:objective_t}, and \ref{tab:objective_ft} show the results of tasks P, T, and PT, respectively. Table \ref{tab:objective_f} shows that the timbre of the source speech was maintained more successfully with DisC than with DisC-NoAux in $F_0$-only conversion. In the same task, we also see that the $F_0$ conversion performance of DisC was better than that of DisC-NoAux. Tables \ref{tab:objective_t} and \ref{tab:objective_ft} show that timbre conversion was more successfully done by DisC than by ACVAE-C, whereas $F_0$ conversion was more successfully done by DisC than by ACVAE-S (especially in timbre-only conversion task). ACVAE-C and -S achieved the best performance in terms of $\Delta F_0$ and the MCD, respectively, however, they could not keep these criteria small at the same time. 

\begin{table}[t]
\caption{MCD from a source speaker and $F_0$ error $\Delta F_0$ in task P. The values before and after $\pm$, respectively, indicate the mean and the standard deviation for all the settings and test samples.}
\label{tab:objective_f}
\scalebox{1}{\begin{tabular}{llll}
 & DisC & DisC-NoAux & (WORLD) \\ \hline
MCD & $6.14 \pm 0.37$ & $6.33 \pm 0.47$ & $4.23 \pm 1.11$ \\ \hline
$\Delta F_0$ & $0.22 \pm 0.10$ & $0.24 \pm 0.11$ & $0.14 \pm 0.12$ \\ \hline
\end{tabular}}
\caption{MCD from a target speaker and $\Delta F_0$ in task T.}
\label{tab:objective_t}
\scalebox{1}{\begin{tabular}{lllll}
 & DisC & DisC-NoAux & ACVAE-C & ACVAE-S \\ \hline
MCD & $7.29 \pm 0.74$ & $7.33 \pm 0.78$ & $7.62 \pm 0.73$ & $6.65 \pm 0.68$ \\ \hline
$\Delta F_0$ & $0.24 \pm 0.14$ & $0.30 \pm 0.20$ & $0.12 \pm 0.11$ & $0.36 \pm 0.18$ \\ \hline
\end{tabular}}
\caption{MCD from a target speaker and $\Delta F_0$ in task PT.}
\label{tab:objective_ft}
\scalebox{1}{\begin{tabular}{llllll}
 & DisC & DisC-NoAux & ACVAE-C & ACVAE-S & (BNE) \\ \hline
MCD & $6.83 \pm 0.63$ & $6.80 \pm 0.63$ & $7.62 \pm 0.74$ & $6.65 \pm 0.68$ & $7.81 \pm 0.67$ \\ \hline
$\Delta F_0$ & $0.13 \pm 0.11$ & $0.14 \pm 0.11$ & $0.12 \pm 0.10$ & $0.14 \pm 0.10$ & $0.19 \pm 0.13$ \\ \hline
\end{tabular}}
\end{table}

\subsection{Subjective evaluation}

\begin{table}[t]
\caption{MOS for sound quality. The values before and after $\pm$, respectively, indicate the mean and the standard deviation for all the listeners, settings, and selected test samples.}
\label{tab:subjective_q}
\scalebox{1}{\begin{tabular}{llllll}
Task & DisC & DisC-NoAux & WORLD & ACVAE-C & ACVAE-S \\ \hline
P & $2.62 \pm 1.05$ & $2.42 \pm 1.12$ & $2.95 \pm 1.17$ & $-$ & $-$ \\ \hline
T & $2.57 \pm 1.15$ & $2.42 \pm 1.19$ & $-$ & $2.24 \pm 0.98$ & $3.55 \pm 0.94$ \\ \hline
PT & $3.45 \pm 0.91$ & $3.77 \pm 0.89$ & $-$ & $2.36 \pm 1.04$ & $3.55 \pm 0.94$ \\ \hline
\end{tabular}}
\caption{Score of ABX test for timbre similarity. Each value indicates similarity with source speakers for task P and that with target speakers  for tasks T and PT.}
\label{tab:subjective_t}
\scalebox{1}{\begin{tabular}{llllll}
Task & DisC & DisC-NoAux & WORLD & ACVAE-C & ACVAE-S \\ \hline
P & $2.03 \pm 0.83$ & $1.85 \pm 0.85$ & $1.34 \pm 0.59$ & $-$ & $-$ \\ \hline
T & $2.06 \pm 1.18$ & $2.06 \pm 1.10$ & $-$ & $1.59 \pm 0.85$ & $3.41 \pm 0.69$ \\ \hline
PT & $3.33 \pm 0.80$ & $3.37 \pm 0.75$ & $-$ & $2.36 \pm 0.88$ & $3.41 \pm 0.69$ \\ \hline
\end{tabular}}
\end{table}

To conduct subjective evaluation, we randomly chose three test samples for each of the quality and timbre evaluations and for each setting (i.e., a combination of $\beta$, a source speaker, and a conversion method in task T and a combination of a source and target speakers and a conversion method in tasks T and PT). 
For timbre evaluation, we concatenated target and converted speech samples with a $1$ s silent interval. We defined the target speech sample as a randomly selected one of the target speaker (i.e., the text content is not necessarily the same as that of the converted one). 
By considering the fact that we use the same result for tasks T and PT with regard to ACVAE-S, each of $15$ listeners was presented $324$ samples in each of quality and timbre evaluations. 

We conducted a mean opinion score (MOS) test for sound quality, where each listener scores the sound quality of each sample by choosing one out of the five options, ``1: Bad,'' ``2: Poor,'' ``3: Fair,'' ``4: Good,'' and ``5: Excellent.'' Timbre evaluation was done based on the ABX test for speaker similarity, where he/she chooses one out of the four options, ``1: Different (sure),'' ``2: Different (not sure),'' ``3: Same (not sure),'' and ``4: Same (sure),'' for each concatenated sample. 

Tables \ref{tab:subjective_q} and \ref{tab:subjective_t}, respectively, show the results of the quality and timbre evaluations. In task P, $F_0$ conversion by WORLD vocoder achieved the best sound quality, however, DisC could maintain the timbre of the source speaker the best. When converting only timbre (i.e., task T), neither DisC nor DisC-NoAux could achieve as high sound quality and timbre maintenance as ACVAE-S. However, by converting $F_0$ pattern and timbre simultaneously (i.e., task PT), they could achieve comparable performance to ACVAE-S. A possible reason for this would be that speaker identity strongly depends on the average fundamental frequency, as also pointed out in \cite{Sambur75}. 


\section{Conclusion}
\label{sec:conclusion}

We developed a new $F_0$-controllable VC method DisC that takes into account discrepancy between the target and generated $F_0$ patterns in the training process. In this paper, we assumed that the same set of speakers appears in training and test data sets. It would be important to extend DisC so that it can be applied to unseen speakers in the test phase. Another future direction would be to take into account other prosody information (e.g., rhythm and stress) as controllable features. 


\clearpage
\bibliographystyle{abbrv}
\bibliography{template}

\end{document}